\documentclass[10pt,a4paper,onecolumn]{article}
\usepackage{marginnote}
\usepackage{graphicx}
\usepackage[rgb]{xcolor}
\usepackage{authblk,etoolbox}
\usepackage{titlesec}
\usepackage{calc}
\usepackage{tikz}
\usepackage[pdfa]{hyperref}
\usepackage{hyperxmp}
\hypersetup{%
    unicode=true,
    pdfapart=3,
    pdfaconformance=B,
    pdftitle={Triumvirate: A Python/C++ package for three-point
clustering measurements},
    pdfauthor={Mike Shengbo Wang, Florian Beutler, Naonori S. Sugiyama},
    pdfpublication={Journal of Open Source Software},
    pdfpublisher={Open Journals},
    pdfissn={2475-9066},
    pdfpubtype={journal},
    pdfvolumenum={},
    pdfissuenum={},
    pdfdoi={10.21105/joss.05571},
    pdfcopyright={Copyright (c) 2023, Mike Shengbo Wang, Florian
Beutler, Naonori S. Sugiyama},
    pdflicenseurl={http://creativecommons.org/licenses/by/4.0/},
    colorlinks=true,
    linkcolor=[rgb]{0.0, 0.5, 1.0},
    citecolor=Blue,
    urlcolor=[rgb]{0.0, 0.5, 1.0},
    breaklinks=true
}
\immediate\pdfobj stream attr{/N 3} file{sRGB.icc}
\pdfcatalog{%
  /OutputIntents [
    <<
      /Type /OutputIntent
      /S /GTS_PDFA1
      /DestOutputProfile \the\pdflastobj\space 0 R
      /OutputConditionIdentifier (sRGB)
      /Info (sRGB)
    >>
  ]
}
\usepackage{caption}
\usepackage{orcidlink}
\usepackage{tcolorbox}
\usepackage{amssymb,amsmath}
\usepackage{ifxetex,ifluatex}
\usepackage{seqsplit}
\usepackage{xstring}

\usepackage{float}
\let\origfigure\figure
\let\endorigfigure\endfigure
\renewenvironment{figure}[1][2] {
    \expandafter\origfigure\expandafter[H]
} {
    \endorigfigure
}

\newlength{\cslhangindent}
\newlength{\csllabelwidth}
\newenvironment{CSLReferences}[2] 
 {
  \setlength{\parindent}{0pt}
  \ifodd #1 \everypar{\setlength{\hangindent}{\cslhangindent}}\ignorespaces\fi
  \ifnum #2 > 0
  \setlength{\parskip}{#2\baselineskip}
  \fi
 }%
 {}
\usepackage{calc}

\usepackage[top=3.5cm, bottom=3cm, right=1.5cm, left=1.0cm,
            headheight=2.2cm, reversemp, includemp, marginparwidth=4.5cm]{geometry}

\titleformat{\section}
  {\normalfont\sffamily\Large\bfseries}
  {}{0pt}{}
\titleformat{\subsection}
  {\normalfont\sffamily\large\bfseries}
  {}{0pt}{}
\titleformat{\subsubsection}
  {\normalfont\sffamily\bfseries}
  {}{0pt}{}
\titleformat*{\paragraph}
  {\sffamily\normalsize}

\usepackage{fancyhdr}
\makeatletter
\let\ps@plain\ps@fancy
\fancyheadoffset[L]{4.5cm}
\fancyfootoffset[L]{4.5cm}

\definecolor{linky}{rgb}{0.0, 0.5, 1.0}

\newtcolorbox{repobox}
   {colback=red, colframe=red!75!black,
     boxrule=0.5pt, arc=2pt, left=6pt, right=6pt, top=3pt, bottom=3pt}

\newcommand{\ExternalLink}{%
   \tikz[x=1.2ex, y=1.2ex, baseline=-0.05ex]{%
       \begin{scope}[x=1ex, y=1ex]
           \clip (-0.1,-0.1)
               --++ (-0, 1.2)
               --++ (0.6, 0)
               --++ (0, -0.6)
               --++ (0.6, 0)
               --++ (0, -1);
           \path[draw,
               line width = 0.5,
               rounded corners=0.5]
               (0,0) rectangle (1,1);
       \end{scope}
       \path[draw, line width = 0.5] (0.5, 0.5)
           -- (1, 1);
       \path[draw, line width = 0.5] (0.6, 1)
           -- (1, 1) -- (1, 0.6);
       }
   }

\patchcmd{\@maketitle}{center}{flushleft}{}{}
\patchcmd{\@maketitle}{center}{flushleft}{}{}
\patchcmd{\@maketitle}{\LARGE}{\LARGE\sffamily}{}{}
\def\maketitle{{%
  
  \AB@maketitle}}
\renewcommand\AB@affilsepx{ \protect\Affilfont}
\renewcommand\AB@affilnote[1]{{\bfseries #1}\hspace{3pt}}
\renewcommand{\affil}[2][]%
   {\newaffiltrue\let\AB@blk@and\AB@pand
      \if\relax#1\relax\def\AB@note{\AB@thenote}\else\def\AB@note{#1}%
        \setcounter{Maxaffil}{0}\fi
        \begingroup
        \let\href=\href@Orig
        \let\protect\@unexpandable@protect
        \def\thanks{\protect\thanks}\def\footnote{\protect\footnote}%
        \@temptokena=\expandafter{\AB@authors}%
        {\def\\{\protect\\\protect\Affilfont}\xdef\AB@temp{#2}}%
         \xdef\AB@authors{\the\@temptokena\AB@las\AB@au@str
         \protect\\[\affilsep]\protect\Affilfont\AB@temp}%
         \gdef\AB@las{}\gdef\AB@au@str{}%
        {\def\\{, \ignorespaces}\xdef\AB@temp{#2}}%
        \@temptokena=\expandafter{\AB@affillist}%
        \xdef\AB@affillist{\the\@temptokena \AB@affilsep
          \AB@affilnote{\AB@note}\protect\Affilfont\AB@temp}%
      \endgroup
       \let\AB@affilsep\AB@affilsepx
}
\makeatother

\renewcommand\Affilfont{\sffamily\small\mdseries}
\ifnum 0\ifxetex 1\fi\ifluatex 1\fi=0 
  \usepackage[T1]{fontenc}
  \usepackage[utf8]{inputenc}

\else 
  \ifxetex
    \usepackage{mathspec}
    \usepackage{fontspec}

  \else
    \usepackage{fontspec}
  \fi
  \defaultfontfeatures{Scale=MatchLowercase}
  \defaultfontfeatures[\sffamily]{Ligatures=TeX}
  \defaultfontfeatures[\rmfamily]{Ligatures=TeX,Scale=1}

\fi
\IfFileExists{upquote.sty}{\usepackage{upquote}}{}
\IfFileExists{microtype.sty}{%
\usepackage{microtype}
\UseMicrotypeSet[protrusion]{basicmath} 
}{}

\PassOptionsToPackage{usenames,dvipsnames}{color} 
\ifLuaTeX
\usepackage[bidi=basic]{babel}
\else
\usepackage[bidi=default]{babel}
\fi
\babelprovide[main,import]{american}

\def\languageshorthands#1{}
\usepackage{longtable,booktabs,array}

\IfFileExists{parskip.sty}{%
\usepackage{parskip}
}{
\setlength{\parindent}{0pt}
\setlength{\parskip}{6pt plus 2pt minus 1pt}
}
\ifx\paragraph\undefined\else
\let\oldparagraph\paragraph
\renewcommand{\paragraph}[1]{\oldparagraph{#1}\mbox{}}
\fi
\ifx\subparagraph\undefined\else
\let\oldsubparagraph\subparagraph
\renewcommand{\subparagraph}[1]{\oldsubparagraph{#1}\mbox{}}
\fi
\ifLuaTeX
  \usepackage{selnolig}  
\fi

\title{Triumvirate: A Python/C++ package for three-point clustering
measurements}

\author[1%
\ensuremath\mathparagraph]{Mike Shengbo Wang%
  \,\orcidlink{0000-0002-2652-4043}\,%
}
\author[1%
]{Florian Beutler%
  \,\orcidlink{0000-0003-0467-5438}\,%
}
\author[2%
]{Naonori S. Sugiyama%
}

\affil[1]{Institute for Astronomy, University of Edinburgh, Royal
Observatory Edinburgh, Blackford Hill, Edinburgh EH9 3HJ, United
Kingdom}
\affil[2]{National Astronomical Observatory of Japan, Mitaka, Tokyo
181-8588, Japan}
\affil[$\mathparagraph$]{Corresponding author}
\date{\vspace{-2.5ex}}

\begin{document}
\maketitle

\marginpar{

  \begin{flushleft}
  \sffamily\small

  {\bfseries DOI:} \href{https://doi.org/10.21105/joss.05571}{\color{linky}{10.21105/joss.05571}}

  \vspace{2mm}
    {\bfseries Software}
  \begin{itemize}
    \setlength\itemsep{0em}
    \item \href{https://github.com/***jounals/***reviews/issues/5571}{\color{linky}{Review}} \ExternalLink
    \item \href{https://github.com/MikeSWang/Triumvirate}{\color{linky}{Repository}} \ExternalLink
    \item \href{https://doi.org/10.5281/zenodo.10072128}{\color{linky}{Archive}} \ExternalLink
  \end{itemize}

  \vspace{2mm}
  
    \par\noindent\hrulefill\par

  \vspace{2mm}

  {\bfseries Editor:} \href{https://github.com/openjournals}{@openjournals} \\
  \vspace{1mm}
    {\bfseries Reviewers:}
  \begin{itemize}
  \setlength\itemsep{0em}
    \item \href{https://github.com/openjournals}{@openjournals}
    \end{itemize}
    \vspace{2mm}
  
    {\bfseries Submitted:} 07 April 2023\\
    {\bfseries Published:} 07 November 2023

  \vspace{2mm}
  {\bfseries License}\\
  Authors of papers retain copyright and release the work under a Creative Commons Attribution 4.0 International License (\href{https://creativecommons.org/licenses/by/4.0/}{\color{linky}{CC BY 4.0}}).

  \end{flushleft}
}

\hypertarget{summary}{%
\section{Summary}\label{summary}}

\texttt{Triumvirate} is a Python/C++ package for measuring the
three-point clustering statistics in large-scale structure (LSS)
cosmological analyses. Given a catalogue of discrete particles (such as
galaxies) with their spatial coordinates, it computes estimators of the
multipoles of the three-point correlation function, also known as the
bispectrum in Fourier space, in the tri-polar spherical harmonic
(TripoSH) decomposition proposed by Sugiyama et al.
(\protect\hyperlink{ref-Sugiyama:2019}{2019}). The objective of
\texttt{Triumvirate} is to provide efficient end-to-end measurement of
clustering statistics which can be fed into downstream galaxy survey
analyses to constrain and test cosmological models. To this end, it
builds upon the original algorithms in the \texttt{hitomi}
code\footnote{\href{https://github.com/naonori/hitomi}{github.com/naonori/hitomi}}
developed by Sugiyama et al.
(\protect\hyperlink{ref-Sugiyama:2018}{2018},
\protect\hyperlink{ref-Sugiyama:2019}{2019}), and supplies a
user-friendly interface with flexible input/output (I/O) of catalogue
data and measurement results, with the built program configurable
through external parameter files and tracked through enhanced logging
and warning/exception handling. For completeness and complementarity,
methods for measuring two-point clustering statistics are also included
in the package.

\hypertarget{statement-of-need}{%
\section{Statement of need}\label{statement-of-need}}

The analysis of higher-order clustering statistics is a key pursuit of
the current and forthcoming generations of large galaxy surveys such as
the Dark Energy Spectroscopic Instrument (DESI)\footnote{\href{https://www.desi.lbl.gov}{desi.lbl.gov}}
(\protect\hyperlink{ref-DESI:2016}{DESI Collaboration et al., 2016}) and
\emph{Euclid}\footnote{\href{https://sci.esa.int/web/euclid/}{sci.esa.int/euclid},
  \href{https://www.euclid-ec.org}{euclid-ec.org}}
(\protect\hyperlink{ref-Euclid:2011}{Euclid Consortium et al., 2011}).
Although the matter density fluctuations in the Universe have been
observed to be almost Gaussian on large scales, primordial
non-Gaussianity in the initial conditions of structure formation and
late-time non-linear gravitational dynamics can both leave potentially
detectable signals in the higher-order moments of the galaxy
distribution on very large and small scales
(\protect\hyperlink{ref-Bernardeau:2002}{Bernardeau et al., 2002}).
Therefore any measurement of the three-point clustering statistics, the
leading non-Gaussian moment, offers a promising probe of both
early-Universe and gravitational physics. In addition, it complements
two-point clustering statistics in constraining cosmological models by
breaking down certain parameter degeneracies
(\protect\hyperlink{ref-Sefusatti:2006}{Sefusatti et al., 2006}).

In contrast to the two-point statistic analysis which has become
standard in recent galaxy surveys (e.g.
\protect\hyperlink{ref-BOSS:2017}{BOSS Collaboration et al., 2017};
\protect\hyperlink{ref-eBOSS:2021}{eBOSS Collaboration et al., 2021}),
three-point clustering statistics have more degrees of freedom and thus
can be compressed in a greater number of ways with different choices of
the coordinate system. For the TripoSH decomposition mentioned above,
there is a need for a computational program that is easy to use,
versatile and suited for the large data sets expected from modern galaxy
surveys, and \texttt{Triumvirate} is designed to meet that demand. More
specifically, it can compute:

\begin{itemize}
\item
  three-point clustering statistics, namely multipoles of the bispectrum
  in Fourier space and of the three-point correlation function (3PCF) in
  configuration space (\protect\hyperlink{ref-Sugiyama:2019}{Sugiyama et
  al., 2019});
\item
  two-point clustering statistics, namely multipoles of the power
  spectrum and two-point correlation function (2PCF)
  (\protect\hyperlink{ref-Sugiyama:2018}{Sugiyama et al., 2018});
\item
  for both three- and two-point statistics, the local plane-parallel
  estimator for a pair of survey-like data and random catalogues, and
  the global plane-parallel estimator for a cubic-box simulation
  (\protect\hyperlink{ref-Feldman:1994}{Feldman et al., 1994};
  \protect\hyperlink{ref-Sugiyama:2019}{Sugiyama et al., 2019};
  \protect\hyperlink{ref-Yamamoto:2006}{Yamamoto et al., 2006});
\item
  for both three- and two-point statistics, the configuration-space
  window function multipoles, which are used to convolve theoretical
  models derived in Fourier space through the Hankel transform
  (\protect\hyperlink{ref-Sugiyama:2019}{Sugiyama et al., 2019};
  \protect\hyperlink{ref-Wilson:2016}{Wilson et al., 2016}).
\end{itemize}

For the global plane-parallel estimators, the simulation box is placed
at the spatial infinity (or equivalently the observer is), so that the
line of sight to each particle can be treated as the same and taken to
be along the \(z\)-axis. For the local plane-parallel estimators, the
observer is placed at the origin in the survey coordinates, and the line
of sight is chosen to point towards one of the particles in a triplet or
pair for three- or two-point clustering measurements respectively.

The geometry of the survey leaves an imprint on the clustering
statistics, where in Fourier space the effect is a convolution with the
survey window function. This convolution mixes different multipoles of
the underlying clustering statistics and the survey window, and the
precise convolution formula (i.e.~the number of multipoles to include in
modelling) needed to achieve a given level of convergence depends on the
precise survey geometry including any sample weights applied. Therefore
the functionality to measure the window function is an integral part of
this program.

These functionalities are essential to cosmological inference pipelines,
and can help validate any analytical covariance matrix predictions
against sample estimates. Since precise covariance matrix estimates
usually require clustering measurements repeated over a large number of
simulated mock catalogues, computational efficiency is an important
objective.

Finally, \texttt{Triumvirate} also enables comparison studies between
alternative compressed statistics of three-point clustering, which may
have different constraining power on different cosmological parameters.
There are existing software packages for some of these alternative
approaches:

\begin{itemize}
\item
  \texttt{pylians}\footnote{\href{https://pylians3.readthedocs.io}{pylians3.readthedocs.io}}
  (\protect\hyperlink{ref-VillaescusaNavarro:2018}{Villaescusa-Navarro,
  2018}) computes the bispectrum with the Scoccimarro estimator
  (\protect\hyperlink{ref-Scoccimarro:2015}{Scoccimarro, 2015}) for
  triangle configurations parametrised by two wavenumbers and the angle
  between the corresponding wavevectors;
\item
  \texttt{nbodykit}\footnote{\href{https://nbodykit.readthedocs.io}{nbodykit.readthedocs.io}}
  (\protect\hyperlink{ref-Hand:2018}{Hand et al., 2018}) computes the
  isotropised 3PCF with a pair-counting algorithm
  (\protect\hyperlink{ref-Slepian:2015}{Slepian \& Eisenstein, 2015}),
  although in principle this can be generalised to anisotropic 3PCF
  (\protect\hyperlink{ref-Slepian:2018}{Slepian \& Eisenstein, 2018}),
  or be implemented using FFTs
  (\protect\hyperlink{ref-Slepian:2016}{Slepian \& Eisenstein, 2016});
\item
  Philcox (\protect\hyperlink{ref-Philcox:2021}{2021})\footnote{\href{https://github.com/oliverphilcox/Spectra-Without-Windows}{github.com/oliverphilcox/Spectra-Without-Windows}}
  advocates a windowless cubic estimator for the bispectrum in the
  Soccimarro decomposition, which can be evaluated using FFTs. However,
  this approach requires the inversion of a Fisher matrix obtained from
  a suite of Monte Carlo realisations.
\end{itemize}

As these programs use a different decomposition of three-point
clustering statistics and focus on either configuration- or
Fourier-space statistics only, \texttt{Triumvirate} fulfills
complementary needs in current galaxy clustering analyses.

\hypertarget{implementation}{%
\section{Implementation}\label{implementation}}

Direct calculation of Fourier modes of density field fluctuations as a
sum over a large number of particles is computationally infeasible, but
the TripoSH estimators can be cast in a form amenable to fast Fourier
transforms (FFTs), which can be utilised to speed up evaluations. In our
numerical scheme, the particles are assigned to regular mesh grids with
appropriate weighting, which is a combination of spherical harmonics and
weights from the input catalogues. Fourier-space fields are obtained by
FFTs over the cubic mesh grids, and clustering statistics are formed by
multiplying the discretely sampled and transformed fields grid-by-grid,
before binning in spherical shells. For the three-point statistics, the
shot noise components are effectively two-point statistics and can be
calculated the same way.

\texttt{Triumvirate} supports mesh assignment schemes up to order 4,
namely the piecewise cubic spline (PCS) scheme, where order \(p\) refers
to the number of grid points a particle is assigned to. Since mesh
assignment convolves the underlying field with a sampling window, the
transformed Fourier-space field should be compensated by dividing out
the sampling window (\protect\hyperlink{ref-Hockney:1988}{Hockney \&
Eastwood, 1988}). For power spectrum measurements only where no inverse
FFT is involved the interlacing technique can be used to reduce the
amount of aliasing (\protect\hyperlink{ref-Sefusatti:2016}{Sefusatti et
al., 2016}), an artifact of discrete Fourier transform where the sampled
Fourier mode at each wavenumber receives contributions from other modes,
with the effect increasingly prominent as the Nyquist wavenumber is
approached; without interlacing, the corrections from Jing
(\protect\hyperlink{ref-Jing:2005}{2005}) (see eq. 20 therein) are
adopted instead. For all clustering statistics, increasing the mesh
assignment order and/or the number of grid cells can help reduce
aliasing (at the expense of speed and memory).

\hypertarget{features}{%
\section{Features}\label{features}}

Besides code refactoring, \texttt{Triumvirate} has many value-added
features in comparison with the predecessor \texttt{himoti}:

\begin{itemize}
\item
  The frontend is written in Python for interactivity and convenience,
  with Cython binding the C++ backend. Although the user will typically
  use the Python interface, the C++ code can also be compiled and
  executed independently.
\item
  Measurement pipelines can be configured through external parameter
  files (in the YAML format for the Python program), cleanly separating
  user inputs from the program itself. Alternatively, measurement
  parameters can be set for individual Python methods without the use of
  a parameter file.
\item
  The reading of catalogue data is implemented via \texttt{astropy.io}
  (\protect\hyperlink{ref-Astropy:2022}{Astropy Collaboration et al.,
  2022}) and \texttt{nbodykit} (\protect\hyperlink{ref-Hand:2018}{Hand
  et al., 2018}), with flexible support for different file formats such
  as text and \texttt{fits} files.
\item
  Numerical algorithms are parallelised with OpenMP, with for-loops over
  catalogue particles and mesh grid cells distributed amongst multiple
  CPU threads.
\item
  Mesh assignment schemes from order \(p = 1\) to \(4\) are supported:
  nearest grid point (NGP), cloud-in-cell (CIC), triangular-shape cloud
  (TSC) and piecewise cubic spline (PCS).
\item
  Interlacing is supported for power spectrum measurements.
\item
  Two normalisation choices are implemented for all clustering
  statistics, one as a sum of catalogue particles
  (\protect\hyperlink{ref-Feldman:1994}{Feldman et al., 1994}, eq.
  2.4.1) and another as a sum over the mesh grid
  (\protect\hyperlink{ref-Sugiyama:2019}{Sugiyama et al., 2019}, eq.
  37).
\item
  A customised logger is provided for runtime tracking, with enhanced
  handling of warnings and exceptions for parameter and data I/O.
\end{itemize}

\hypertarget{performance}{%
\section{Performance}\label{performance}}

When a large number of grid cells, \(N_\mathrm{mesh}\), are used to
sample the density fields from a catalogue of \(N_\mathrm{part}\)
particles on a mesh with \(N_\mathrm{mesh} \gg N_\mathrm{part}\), the
dominant operations are FFTs with complexity
\(\mathcal{O}\left({N_\mathrm{mesh} \ln N_\mathrm{mesh}}\right)\).
Therefore the complexity for three-point clustering measurements is
\(\mathcal{O}\left({N_\mathrm{bin}^2 N_\mathrm{mesh} \ln N_\mathrm{mesh}}\right)\),
where \(N_\mathrm{bin}\) is the number of coordinate bins.

It is worth noting that in \texttt{Triumvirate}, the spherical harmonic
weights are applied to individual particles rather than the mesh grids.
This should result in more accurate results at the expense of memory
usage, as multiple meshes need to be stored for spherical harmonics of
different degrees and orders. We estimate the minimum memory usage for
bispectrum measurements to be \(11 M\) and \(9 M\) respectively for
local and global plane-parallel estimators, where
\(M = 16 N_\mathrm{mesh}\) bytes (roughly
\(1.5\times10^{-8} N_\mathrm{mesh}\) gibibytes\footnote{Note that 1
  gibibyte (GiB) is \(2^{30}\) bytes, as opposed to 1 gigabyte (GB)
  which is \(10^9\) bytes. GiB is the preferred unit by job schedulers
  such as Slurm for computer clusters.}); for local and global
plane-parallel 3PCF estimators, the figures are \(10 M\) and \(9 M\)
respectively.

In the table below, we show the wall time and peak memory usage for
bispectrum and 3PCF measurements of a few select multipoles and grid
numbers with \(N_\mathrm{bin} = 20\), using a single core on one AMD
EPYC 7H12 processor with base frequency 2.60 GHz. With multithreading
enabled, the run time is reduced (see the last column in the table).
Here ‘lpp’ and ‘gpp’ denote local and global plane-parallel
approximations respectively. For the global plane-parallel estimates,
the catalogue used is a cubic box containing
\(N_\mathrm{part} = 8 \times 10^6\) particles; for the local
plane-parallel estimates, the data and random catalogues contain
\(N_\mathrm{part} = 6.6 \times 10^5\) and \(1.3 \times 10^7\) particles
respectively. Since both the bispectrum and 3PCF are computed with FFTs,
the computation time and memory usage for them are roughly the same,
with minor differences due to the slightly different number of mesh
grids needed for evaluation.

\begin{longtable}[]{@{}
  >{\raggedright\arraybackslash}p{(\columnwidth - 8\tabcolsep) * \real{0.2109}}
  >{\raggedleft\arraybackslash}p{(\columnwidth - 8\tabcolsep) * \real{0.1905}}
  >{\raggedleft\arraybackslash}p{(\columnwidth - 8\tabcolsep) * \real{0.1905}}
  >{\raggedleft\arraybackslash}p{(\columnwidth - 8\tabcolsep) * \real{0.1905}}
  >{\raggedleft\arraybackslash}p{(\columnwidth - 8\tabcolsep) * \real{0.2177}}@{}}
\toprule\noalign{}
\begin{minipage}[b]{\linewidth}\raggedright
Multipole/\(N_\mathrm{mesh}\)
\end{minipage} & \begin{minipage}[b]{\linewidth}\raggedleft
\(128^3\)
\end{minipage} & \begin{minipage}[b]{\linewidth}\raggedleft
\(256^3\)
\end{minipage} & \begin{minipage}[b]{\linewidth}\raggedleft
\(512^3\)
\end{minipage} & \begin{minipage}[b]{\linewidth}\raggedleft
\(512^3\) (32 threads)
\end{minipage} \\
\midrule\noalign{}
\endhead
\bottomrule\noalign{}
\endlastfoot
\(B_{000}^{\mathrm{(lpp)}}\) & 96 s, 1.8 GiB & 215 s, 4.4 GiB & 1247 s,
25 GiB & 85 s, 25 GiB \\
\(B_{000}^{\mathrm{(gpp)}}\) & 42 s, 0.7 GiB & 172 s, 2.9 GiB & 1185 s,
21 GiB & 59 s, 21 GiB \\
\(B_{202}^{\mathrm{(lpp)}}\) & 320 s, 1.8 GiB & 1030 s, 4.4 GiB & 6449
s, 25 GiB & 267 s, 25 GiB \\
\(B_{202}^{\mathrm{(gpp)}}\) & 42 s, 0.7 GiB & 176 s, 2.9 GiB & 1187 s,
21 GiB & 60 s, 21 GiB \\
\(\zeta_{000}^{\mathrm{(lpp)}}\) & 90 s, 1.8 GiB & 211 s, 4.4 GiB & 1403
s, 21 GiB & 83 s, 21 GiB \\
\(\zeta_{000}^{\mathrm{(gpp)}}\) & 43 s, 0.7 GiB & 178 s, 2.9 GiB & 1226
s, 19 GiB & 55 s, 19 GiB \\
\(\zeta_{202}^{\mathrm{(lpp)}}\) & 267 s, 1.8 GiB & 964 s, 4.4 GiB &
6377 s, 21 GiB & 266 s, 21 GiB \\
\(\zeta_{202}^{\mathrm{(gpp)}}\) & 43 s, 0.7 GiB & 177 s, 2.9 GiB & 1241
s, 19 GiB & 57 s, 19 GiB \\
\end{longtable}

\hypertarget{future-work}{%
\section{Future work}\label{future-work}}

\texttt{Triumvirate} will be routinely maintained and updated depending
on user feedback. One extension of interest is the inclusion of other
three-point clustering estimators with different coordinate systems and
compression choices, and the functionality to transform between them.
The ability to measure clustering statistics from a density field
already sampled on a mesh grid may also be useful. In addition, porting
the code to graphic processing units (GPUs) can bring further
parallelisation that can enhance the performance of the code.

\hypertarget{acknowledgements}{%
\section{Acknowledgements}\label{acknowledgements}}

This project has received funding from the European Research Council
(ERC) under the European Union’s Horizon 2020 research and innovation
programme (grant agreement 853291). FB is a Royal Society University
Research Fellow.

We thank the reviewers for their valuable feedback and suggestions,
which have improved the functionality and documentation of this code.

For the purpose of open access, the author has applied a
Creative Commons Attribution (CC BY) licence to any Author Accepted
Manuscript version arising from this submission.  

\hypertarget{references}{%
\section*{References}\label{references}}
\addcontentsline{toc}{section}{References}

\hypertarget{refs}{}
\begin{CSLReferences}{1}{0}
\leavevmode\vadjust pre{\hypertarget{ref-Astropy:2022}{}}%
Astropy Collaboration, Price-Whelan, A. M., Lim, P. L., Earl, N.,
Starkman, N., Bradley, L., Shupe, D. L., Patil, A. A., Corrales, L.,
Brasseur, C. E., Nöthe, M., Donath, A., Tollerud, E., Morris, B. M.,
Ginsburg, A., Vaher, E., Weaver, B. A., Tocknell, J., Jamieson, W., …
Astropy Project Contributors. (2022). {The Astropy Project: Sustaining
and Growing a Community-oriented Open-source Project and the Latest
Major Release (v5.0) of the Core Package}. \emph{Astrophys.~J.},
\emph{935}(2), 167. \url{https://doi.org/10.3847/1538-4357/ac7c74}

\leavevmode\vadjust pre{\hypertarget{ref-Bernardeau:2002}{}}%
Bernardeau, F., Colombi, S., Gaztañaga, E., \& Scoccimarro, R. (2002).
Large-scale structure of the universe and cosmological perturbation
theory. \emph{Phys.~Rep.}, \emph{367}, 1–248.
\url{https://doi.org/10.1016/s0370-1573(02)00135-7}

\leavevmode\vadjust pre{\hypertarget{ref-BOSS:2017}{}}%
BOSS Collaboration, Alam, S., Ata, M., Bailey, S., Beutler, F., Bizyaev,
D., Blazek, J. A., Bolton, A. S., Brownstein, J. R., Burden, A., Chuang,
C.-H., Comparat, J., Cuesta, A. J., Dawson, K. S., Eisenstein, D. J.,
Escoffier, S., Gil-Mar\'{i}n, H., Grieb, J. N., Hand, N., … Zhao, G.-B.
(2017). The clustering of galaxies in the completed {SDSS-III Baryon
Oscillation Spectroscopic Survey}: Cosmological analysis of the {DR}12
galaxy sample. \emph{Mon.~Not.~R.~Astron.~Soc.}, \emph{470}(3),
2617–2652. \url{https://doi.org/10.1093/mnras/stx721}

\leavevmode\vadjust pre{\hypertarget{ref-DESI:2016}{}}%
DESI Collaboration, Aghamousa, A., Aguilar, J., Ahlen, S., Alam, S.,
Allen, L. E., Allende Prieto, C., Annis, J., Bailey, S., Balland, C.,
Ballester, O., Baltay, C., Beaufore, L., Bebek, C., Beers, T. C., Bell,
E. F., Bernal, J. L., Besuner, R., Beutler, F., … Zu, Y. (2016). {The
DESI Experiment Part I: Science, Targeting, and Survey Design}.
\emph{arXiv e-Prints}. \url{https://doi.org/10.48550/arXiv.1611.00036}

\leavevmode\vadjust pre{\hypertarget{ref-eBOSS:2021}{}}%
eBOSS Collaboration, Alam, S., Aubert, M., Avila, S., Balland, C.,
Bautista, J. E., Bershady, M. A., Bizyaev, D., Blanton, M. R., Bolton,
A. S., Bovy, J., Brinkmann, J., Brownstein, J. R., Burtin, E.,
Chabanier, S., Chapman, M. J., Choi, P. D., Chuang, C.-H., Comparat, J.,
… Zheng, Z. (2021). Completed {SDSS-IV extended Baryon Oscillation
Spectroscopic Survey}: {Cosmological} implications from two decades of
spectroscopic surveys at the {Apache Point Observatory}.
\emph{Phys.~Rev.~D}, \emph{103}, 083533.
\url{https://doi.org/10.1103/PhysRevD.103.083533}

\leavevmode\vadjust pre{\hypertarget{ref-Euclid:2011}{}}%
Euclid Consortium, Laureijs, R., Amiaux, J., Arduini, S., Auguères,
J.-L., Brinchmann, J., Cole, R., Cropper, M., Dabin, C., Duvet, L.,
Ealet, A., Garilli, B., Gondoin, P., Guzzo, L., Hoar, J., Hoekstra, H.,
Holmes, R., Kitching, T., Maciaszek, T., … Zucca, E. (2011). {Euclid
Definition Study Report}. \emph{arXiv e-Prints}.
\url{https://doi.org/10.48550/arXiv.1110.3193}

\leavevmode\vadjust pre{\hypertarget{ref-Feldman:1994}{}}%
Feldman, H. A., Kaiser, N., \& Peacock, J. A. (1994). {Power-Spectrum
Analysis of Three-dimensional Redshift Surveys}. \emph{Astrophys.~J.},
\emph{426}, 23. \url{https://doi.org/10.1086/174036}

\leavevmode\vadjust pre{\hypertarget{ref-Hand:2018}{}}%
Hand, N., Feng, Y., Beutler, F., Li, Y., Modi, C., Seljak, U., \&
Slepian, Z. (2018). {nbodykit: An Open-source, Massively Parallel
Toolkit for Large-scale Structure}. \emph{Astron.~J.}, \emph{156}(4),
160. \url{https://doi.org/10.3847/1538-3881/aadae0}

\leavevmode\vadjust pre{\hypertarget{ref-Hockney:1988}{}}%
Hockney, R. W., \& Eastwood, J. W. (1988). \emph{{Computer Simulation
Using Particles}} (1st ed.). CRC~Press.
\url{https://doi.org/10.1201/9780367806934}

\leavevmode\vadjust pre{\hypertarget{ref-Jing:2005}{}}%
Jing, Y. P. (2005). {Correcting for the Alias Effect When Measuring the
Power Spectrum Using a Fast Fourier Transform}. \emph{Astrophys.~J.},
\emph{620}(2), 559. \url{https://doi.org/10.1086/427087}

\leavevmode\vadjust pre{\hypertarget{ref-Philcox:2021}{}}%
Philcox, O. H. E. (2021). Cosmology without window functions. {II}.
Cubic estimators for the galaxy bispectrum. \emph{Phys.~Rev.~D},
\emph{104}(12), 123529.
\url{https://doi.org/10.1103/physrevd.104.123529}

\leavevmode\vadjust pre{\hypertarget{ref-Scoccimarro:2015}{}}%
Scoccimarro, R. (2015). Fast estimators for redshift-space clustering.
\emph{Phys.~Rev.~D}, \emph{92}, 083532.
\url{https://doi.org/10.1103/PhysRevD.92.083532}

\leavevmode\vadjust pre{\hypertarget{ref-Sefusatti:2006}{}}%
Sefusatti, E., Crocce, M., Pueblas, S., \& Scoccimarro, R. (2006).
Cosmology and the bispectrum. \emph{Phys.~Rev.~D}, \emph{74}, 023522.
\url{https://doi.org/10.1103/PhysRevD.74.023522}

\leavevmode\vadjust pre{\hypertarget{ref-Sefusatti:2016}{}}%
Sefusatti, E., Crocce, M., Scoccimarro, R., \& Couchman, H. M. P.
(2016). Accurate estimators of correlation functions in {Fourier} space.
\emph{Mon.~Not.~R.~Astron.~Soc.}, \emph{460}(4), 3624–3636.
\url{https://doi.org/10.1093/mnras/stw1229}

\leavevmode\vadjust pre{\hypertarget{ref-Slepian:2015}{}}%
Slepian, Z., \& Eisenstein, D. J. (2015). Computing the three-point
correlation function of galaxies in \(\mathcal{O}(N^2)\) time.
\emph{Mon.~Not.~R.~Astron.~Soc.}, \emph{454}(4), 4142–4158.
\url{https://doi.org/10.1093/mnras/stv2119}

\leavevmode\vadjust pre{\hypertarget{ref-Slepian:2016}{}}%
Slepian, Z., \& Eisenstein, D. J. (2016). Accelerating the two-point and
three-point galaxy correlation functions using fourier transforms.
\emph{Mon.~Not.~R.~Astron.~Soc.}, \emph{455}(1), L31–L35.
\url{https://doi.org/10.1093/mnrasl/slv133}

\leavevmode\vadjust pre{\hypertarget{ref-Slepian:2018}{}}%
Slepian, Z., \& Eisenstein, D. J. (2018). A practical computational
method for the anisotropic redshift-space three-point correlation
function. \emph{Mon.~Not.~R.~Astron.~Soc.}, \emph{478}(2), 1468–1483.
\url{https://doi.org/10.1093/mnras/sty1063}

\leavevmode\vadjust pre{\hypertarget{ref-Sugiyama:2019}{}}%
Sugiyama, N. S., Saito, S., Beutler, F., \& Seo, H.-J. (2019). A
complete {FFT}-based decomposition formalism for the redshift-space
bispectrum. \emph{Mon.~Not.~R.~Astron.~Soc.}, \emph{484}(1), 364–384.
\url{https://doi.org/10.1093/mnras/sty3249}

\leavevmode\vadjust pre{\hypertarget{ref-Sugiyama:2018}{}}%
Sugiyama, N. S., Shiraishi, M., \& Okumura, T. (2018). {Limits on
statistical anisotropy from {BOSS DR12} galaxies using bipolar spherical
harmonics}. \emph{Mon.~Not.~R.~Astron.~Soc.}, \emph{473}(2), 2737–2752.
\url{https://doi.org/10.1093/mnras/stx2333}

\leavevmode\vadjust pre{\hypertarget{ref-VillaescusaNavarro:2018}{}}%
Villaescusa-Navarro, F. (2018). \emph{Pylians: {Python} libraries for
the analysis of numerical simulations}. {Astrophysics Source Code
Library} {[}ascl:1811.008{]}.

\leavevmode\vadjust pre{\hypertarget{ref-Wilson:2016}{}}%
Wilson, M. J., Peacock, J. A., Taylor, A. N., \& de la Torre, S. (2016).
Rapid modelling of the redshift-space power spectrum multipoles for a
masked density field. \emph{Mon.~Not.~R.~Astron.~Soc.}, \emph{464}(3),
3121–3130. \url{https://doi.org/10.1093/mnras/stw2576}

\leavevmode\vadjust pre{\hypertarget{ref-Yamamoto:2006}{}}%
Yamamoto, K., Nakamichi, M., Kamino, A., Bassett, B. A., \& Nishioka, H.
(2006). {A Measurement of the Quadrupole Power Spectrum in the
Clustering of the 2dF QSO Survey}. \emph{Publ. Astron. Soc. Jpn.},
\emph{58}(1), 93–102. \url{https://doi.org/10.1093/pasj/58.1.93}

\end{CSLReferences}

\end{document}